\documentclass[twocolumn,showpacs,preprintnumbers,
prb,superscriptaddress]{revtex4}%
\usepackage{amssymb}
\usepackage{amsmath}
\usepackage{graphicx}
\usepackage{dcolumn}
\usepackage{bm}
\usepackage{amsfonts}%
\providecommand{\U}[1]{\protect\rule{.1in}{.1in}}

\begin{document}
\title{Green function theory of orbital magnetic moment of interacting electrons in solids}
\author{F. Aryasetiawan}
\affiliation{Department of Physics, Division of Mathematical Physics, Lund University,
S\"{o}lvegatan 14A, 223 62 Lund, Sweden}
\author{K. Karlsson}
\affiliation{Department of Engineering Sciences, University of Sk\"{o}vde, SE-541 28
Sk\"{o}vde, Sweden}
\author{T. Miyake}
\affiliation{CD-FMat, AIST, Tsukuba 305-8568, Japan}
\affiliation{ESICMM, National Institute for Materials Science, Tsukuba 305-0047, Japan}
\date{\today }

\begin{abstract}
A general formula for the orbital magnetic moment of interacting electrons in
solids is derived using the many-electron Green function method. The formula
factorizes into two parts, a part that contains the information about the
one-particle band structure of the system and a part that contains the effects
of exchange and correlations carried by the Green function. The derived
formula provides a convenient means of including the effects of exchange and
correlations beyond the commonly used local density approximation of density
functional theory.

\end{abstract}

\pacs{71.20.-b, 71.27.+a}
\maketitle

The magnetic moment of an atom is composed of two components: one arises from
the spins of the electrons and another from the orbital motion of the
electrons around the nucleus. When atoms are assembled together to form a
crystal, these magnetic moments originating from the individual atoms survive.
Until recently, it was thought that calculating the magnetic moment of a
crystal amounts to calculating the local magnetic moments of the constituent
atoms. Some years ago, however, it was discovered that in addition to the
contribution from the local atomic orbital moments there is a contribution
emerging from the itinerant motion of the electrons on the surface of the
crystal \cite{thonhauser2005, xiao2005}. In the thermodynamic limit when the
system size becomes very large, this contribution does not depend on the
geometry of the surface and can be expressed in terms of bulk quantities.

Although in many cases, the magnetic moment is dominated by the contribution
from the electron spins, in some cases the contribution from the orbital
moment can be significant. It is therefore important to have a rigorous
formulation for calculating the orbital moment contribution to the total
magnetic moment. It has been recognized for a long time that calculating the
orbital magnetic moment of a crystal is a nontrivial task due to the infinite
nature of a crystal that renders the position operator entering the expression
for the orbital moment ill-defined \cite{resta1998,resta2005}. In this
respect, the problem is akin to the calculation of the polarization of a
crystal which was solved by King-Smith and Vanderbilt
\cite{king-smith1993,vanderbilt1993} in the early nineties by utilizing
Wannier orbitals and the Berry-phase approach \cite{berry1984}. A similar
approach was proposed by Resta \emph{et al} \cite{resta1998,resta2005} for
calculating the orbital magnetic moment in crystals and led to the finding of
the possible presence of an additional itinerant contribution which was put on
a firm theoretical basis in a later work \cite{thonhauser2005}. Independently
and parallel to the Wannier-orbital approach, a semiclassical method was
proposed by Xiao \emph{et al} \cite{xiao2005} and later a linear-response
technique by Shi \emph{et al }\cite{shi2007} to derive a general expression
for the orbital magnetic moment of a crystal. A mathematically elegant
derivation was also presented very recently by Bianco and Resta
\cite{bianco2013}.

These pioneering works have opened up a way for computing the orbital magnetic
moment of a crystal from realistic band structure calculations
\cite{ceresoli10,lopez12}. The focus, however, has been on non-interacting
systems represented by one-particle Bloch states and the corresponding
eigenvalues usually obtained from the Kohn-Sham theory of density functional
theory. For interacting systems, the subject is still new and unexplored. An
extension to interacting systems was proposed within the framework of the
current and spin density functional theory which provides an appealing
analogue to the non-interacting formulation. While current and spin density
functional theory is formally exact it is common practice to use the local
density approximation for the exchange-correlation potential. However, it is
hard to improve upon the local density approximation (LDA) in a systematic fashion.

In this letter, we present a general alternative method for calculating the
orbital magnetic moment of interacting many-electron systems in a periodic
lattice based on the many-body Green function technique. Green function method
is rather well established and allows for a more systematic improvement
compared with the density-functional based approaches. Indeed, in the last
couple of decades there has been a number of important breakthroughs in
applying it to calculate the electronic structure and spectrocopic properties
of realistic materials. The \emph{GW} approximation
\cite{hedin65,aryasetiawan98,aulbur00,onida02} and the dynamical mean-field
theory (DMFT) \cite{georges96} in combination with the LDA (LDA+DMFT) are
prime examples of Green-function based theories successfully applied to real
materials. These two methods provide effective tools for describing the
electronic structure of materials from weakly to strongly correlated.

The orbital moment of a many-electron system is given by%

\begin{equation}
\mathbf{M}_{\text{orb}}=\left\langle \Psi|\mathbf{\hat{L}}|\Psi\right\rangle
\end{equation}
where $\Psi$ is the many-electron ground state and the angular momentum
operator is given by%

\begin{equation}
\mathbf{\hat{L}}=\int dr\hat{\psi}^{\dag}(r)\mathbf{L}(r)\hat{\psi
}(r),~~\mathbf{L}=\mathbf{r\times v}. \label{L}%
\end{equation}
We use the notation $r=(\mathbf{r,}\sigma\mathbf{)}$ and work in atomic unit
($\hslash=m=e=1$) throughout the derivation. The magnetic moment is obtained
by multiplying the orbital moment by a factor $-e/2c$ (Gaussian units). The
velocity operator may be expressed as a commutator%

\begin{equation}
\mathbf{v}=i[H,\mathbf{r}]. \label{Hr}%
\end{equation}
$H$ is a one-particle Hamiltonian whose eigenfunctions given by%

\begin{equation}
H\psi_{\mathbf{k}n}=E_{\mathbf{k}n}\psi_{\mathbf{k}n}%
\end{equation}
may be chosen to define the field operator $\hat{\psi}$ in Eq. (\ref{L}). For
example, $H$ is chosen in practice to be the Kohn-Sham Hamiltonian. We can
express $M_{\text{orb}}$ in terms of many-electron Green function as follows:%

\begin{equation}
\mathbf{M}_{\text{orb}}=-i\lim_{r^{\prime}\rightarrow r}\int dr\mathbf{L}%
(r^{\prime})G^{+}(r^{\prime},r),
\end{equation}

\begin{equation}
G^{+}(r^{\prime},r)=i\left\langle \Psi|\hat{\psi}^{\dag}(rt^{+})\hat{\psi
}(r^{\prime}t)|\Psi\right\rangle .
\end{equation}

We may expand the Green function in maximally localized Wannier orbitals
\cite{marzari97, souza01} $\{\varphi_{\mathbf{R}n}\}$ and to simplify writing
we use a notation in which repeated indices are summed:%

\begin{equation}
G^{+}(r^{\prime},r)=\varphi_{\mathbf{R}^{\prime}n^{\prime}}(r^{\prime
})G_{\mathbf{R}^{\prime}n^{\prime},\mathbf{R}n}^{+}\varphi_{\mathbf{R}n}%
^{\ast}(r)
\end{equation}
where%

\begin{align}
\varphi_{\mathbf{R}n}(r)  &  =\frac{1}{N}e^{-i\mathbf{k\cdot R}}\tilde{\psi
}_{\mathbf{k}n}(r),\text{ \ \ }\tilde{\psi}_{\mathbf{k}n}%
(r)=e^{i\mathbf{k\cdot R}}\varphi_{\mathbf{R}n}(r),\\
\tilde{\psi}_{\mathbf{k}n}(r)  &  =\psi_{\mathbf{k}m}S_{mn}(\mathbf{k}).
\end{align}
With the above definitions, $\varphi_{\mathbf{R}n}$ is normalized over the
entire space whereas $\tilde{\psi}_{\mathbf{k}n}$ is normalized over the unit
cell. The transformation matrix $S_{mn}$ ensures that a given band labelled by
$n$ is smooth in the Wannier basis \cite{marzari97, souza01}. The Wannier
orbitals are used as a tool but the final formula for the orbital magnetic
moments is expressed entirely in terms of Bloch states. In this work, we
restrict ourselves to systems with zero Chern number since the above
construction of Wannier orbitals may not be valid for systems with nonzero
Chern number.

Using the Wannier expansion, the total orbital moment can be split into a
local and an itinerant term:%

\begin{equation}
\mathbf{M}^{L}=-iG_{\mathbf{R}^{\prime}n^{\prime},\mathbf{R}n}^{+}\int
dr\varphi_{\mathbf{R}n}^{\ast}(r)[(\mathbf{r-R})\times\mathbf{v]}%
\varphi_{\mathbf{R}^{\prime}n^{\prime}}(r), \label{ML}%
\end{equation}

\begin{equation}
\mathbf{M}^{I}=-iG_{\mathbf{R}^{\prime}n^{\prime},\mathbf{R}n}^{+}\int
dr\varphi_{\mathbf{R}n}^{\ast}(r)(\mathbf{R}\times\mathbf{v)}\varphi
_{\mathbf{R}^{\prime}n^{\prime}}(r). \label{MI}%
\end{equation}
In the thermodynamic limit, when the system is made increasingly large, each
atom labelled by $\mathbf{R}$ contributes equally to $\mathbf{M}^{L}$ and we
may set $\mathbf{R}=\mathbf{0}$ and multiply by the number of atoms. We are
assuming for simplicity that there is one atom per unit cell but extension to
many atoms per unit cell is quite straightforward and the final formula is not
affected by this assumption. The local orbital moment per unit volume is then%

\begin{equation}
\mathbf{m}^{L}=-\frac{i}{\Omega}G_{\mathbf{R}^{\prime}n^{\prime},\mathbf{0}%
n}^{+}\int dr\varphi_{\mathbf{0}n}^{\ast}(r)\left(  \mathbf{r\times v}\right)
\varphi_{\mathbf{R}^{\prime}n^{\prime}}(r). \label{morbL1}%
\end{equation}
Eq. (\ref{morbL1}) has a clear physical interpretation: For $\mathbf{R}%
^{\prime}=\mathbf{0}$ it corresponds to the rotation of the Wannier orbital
around the origin. For $\mathbf{R}^{\prime}\neq\mathbf{0}$ the Green function
$G_{\mathbf{R}^{\prime}n^{\prime},\mathbf{0}n}^{+}$ is the probability
amplitude for an electron at a Wannier orbital $n$ located at the origin to
hop to a Wannier orbital $n^{\prime}$ located at $\mathbf{R}^{\prime}$.
Multiplying it with the matrix element $\left\langle \varphi_{\mathbf{0}%
n}|\mathbf{r\times v}|\varphi_{\mathbf{R}^{\prime}n^{\prime}}\right\rangle $
gives the contribution to the orbital moment from that electron that hops to
$\varphi_{\mathbf{R}^{\prime}n^{\prime}}$. For a non-interacting system, the
$\mathbf{R=0}$ term in (\ref{morbL1}) corresponds to Eq. (16) of Thonhauser
\cite{thonhauser2011} and for an occupied band, it can be shown, assuming
$S=1$ and using%
\begin{equation}
G_{n^{\prime}n}^{+}(\mathbf{q})=G_{\mathbf{R}^{\prime}n^{\prime},\mathbf{0}%
n}^{+}e^{-i\mathbf{q\cdot R}^{\prime}}=in_{F}(E_{\mathbf{q}n})\delta
_{n^{\prime}n} \label{G0}%
\end{equation}
where $n_{F}$ is the Fermi function, that the $\mathbf{R\neq0}$ terms in Eq.
(\ref{morbL1}) vanish but they may be finite for metallic systems.

The itinerant contribution $\mathbf{M}^{I}$ for $\mathbf{R=R}^{\prime}$ and
$n=n^{\prime}$ is interpreted as arising from the center-of-mass motion of the
Wannier orbitals $\varphi_{\mathbf{R}n}$ on the surface. The $\mathbf{R\neq
R}^{\prime}$ and $n\neq n^{\prime}$ terms describe an additional itinerant
contribution from the possibility of hopping from $\varphi_{\mathbf{R}n}$ to
$\varphi_{\mathbf{R}^{\prime}n^{\prime}}$.

We now express $\mathbf{m}^{L}$ in terms of Bloch states. Using the commutator
in Eq. (\ref{Hr}) and the Levi-Civita symbol $\epsilon_{ijk}$ we find%

\begin{equation}
m_{i}^{L}=\frac{1}{\Omega}G_{\mathbf{R}^{\prime}n^{\prime},\mathbf{0}n}%
^{+}\epsilon_{ijk}\int dr\varphi_{\mathbf{0}n}^{\ast}(r)r_{j}Hr_{k}%
\varphi_{\mathbf{R}^{\prime}n^{\prime}}(r) \label{miL}%
\end{equation}
where the other term arising from the commutator is zero since $\sum
_{jk}\epsilon_{ijk}r_{j}r_{k}=0$. Using the identity%

\begin{equation}
r_{k}\varphi_{\mathbf{R}^{\prime}n^{\prime}}(r)=i\frac{\Omega}{(2\pi)^{3}}\int
d^{3}q^{\prime}e^{i\mathbf{q}^{\prime}\mathbf{\cdot r}}\frac{\partial
}{\partial q_{k}^{\prime}}[e^{-i\mathbf{q}^{\prime}\mathbf{\cdot R}^{\prime}%
}\tilde{u}_{\mathbf{q}^{\prime}n^{\prime}}(r\mathbf{)]} \label{rphi}%
\end{equation}
where $\tilde{u}_{\mathbf{q}n}=\exp(-i\mathbf{q\cdot r)}\tilde{\psi
}_{\mathbf{q}n}$ is the periodic part of the Bloch function and $\Omega$ is
the unit cell volume, we obtain%

\begin{align}
m_{i}^{L}  &  =\epsilon_{ijk}\int_{BZ}\frac{d^{3}q}{(2\pi)^{3}}\left\{
G_{n^{\prime}n}^{+}(\mathbf{q})\left\langle \frac{\partial\tilde
{u}_{\mathbf{q}n}}{\partial q_{j}}\right\vert H(\mathbf{q})\left\vert
\frac{\partial\tilde{u}_{\mathbf{q}n^{\prime}}}{\partial q_{k}}\right\rangle
\right. \nonumber\\
&  +\left.  \frac{\partial G_{n^{\prime}n}^{+}(\mathbf{q})}{\partial q_{k}%
}\left\langle \frac{\partial\tilde{u}_{\mathbf{q}n}}{\partial q_{j}%
}|u_{\mathbf{q}m^{\prime}}\right\rangle E_{\mathbf{q}m^{\prime}}S_{m^{\prime
}n^{\prime}}(\mathbf{q)}\right\}  \mathbf{.} \label{mloc}%
\end{align}
where the integration over the crystal momentum space has been reduced to the
Brillouin zone (BZ) and%

\begin{equation}
H(\mathbf{q})=e^{-i\mathbf{q\cdot r}}H(r)e^{i\mathbf{q\cdot r}},
\end{equation}

\begin{equation}
G_{n^{\prime}n}^{+}(\mathbf{q})\doteqdot\sum_{\mathbf{R}}G_{\mathbf{R}%
n^{\prime},\mathbf{0}n}^{+}e^{-i\mathbf{q\cdot R}}. \label{Gq}%
\end{equation}

We now consider the itinerant term in (\ref{MI}). We need to evaluate
$\left\langle \varphi_{\mathbf{R}n}|\mathbf{v}|\varphi_{\mathbf{R}^{\prime
}n^{\prime}}\right\rangle $, which can be done by inserting a complete set of
states in between $H$ and $\mathbf{r}$:%

\begin{align}
&  \left\langle \varphi_{\mathbf{R}n}|\mathbf{v}|\varphi_{\mathbf{R}^{\prime
}n^{\prime}}\right\rangle \nonumber\\
&  =i[\left\langle \varphi_{\mathbf{R}n}|H\left\vert \varphi_{\mathbf{R}%
^{\prime\prime}m}\right\rangle \left\langle \varphi_{\mathbf{R}^{\prime\prime
}m}\right\vert \mathbf{r}|\varphi_{\mathbf{R}^{\prime}n^{\prime}}\right\rangle
\nonumber\\
&  -\left\langle \varphi_{\mathbf{R}n}|\mathbf{r}\left\vert \varphi
_{\mathbf{R}^{\prime\prime}m}\right\rangle \left\langle \varphi_{\mathbf{R}%
^{\prime\prime}m}\right\vert H|\varphi_{\mathbf{R}^{\prime}n^{\prime}%
}\right\rangle ].
\end{align}
We may choose $\mathbf{R}^{\prime\prime}$ as the origin of the coordinate and
measure all vectors with respect to $\mathbf{R}^{\prime\prime}$. Since the
sums over $\mathbf{R}$ and $\mathbf{R}^{\prime}$ run over all lattice points,
we may relabel $\mathbf{R-R}^{\prime\prime}\rightarrow\mathbf{R}$ and
$\mathbf{R}^{\prime}-\mathbf{R}^{\prime\prime}\rightarrow\mathbf{R}^{\prime}$.
Both $H$ and $G$ are not affected by the relabelling since they are periodic.
We then find%

\begin{align}
\mathbf{M}^{I}  &  =-iG_{\mathbf{R}^{\prime}n^{\prime},\mathbf{R}n}%
^{+}(\mathbf{R+R}^{\prime\prime})\times\nonumber\\
&  i\left[  \left\langle \varphi_{\mathbf{R}n}|H\left\vert \varphi
_{\mathbf{0}m}\right\rangle \left\langle \varphi_{\mathbf{0}m}\right\vert
\mathbf{r+R}^{\prime\prime}|\varphi_{\mathbf{R}^{\prime}n^{\prime}%
}\right\rangle \right. \nonumber\\
&  -\left.  \left\langle \varphi_{\mathbf{R}n}|\mathbf{r+R}^{\prime\prime
}\left\vert \varphi_{\mathbf{0}m}\right\rangle \left\langle \varphi
_{\mathbf{0}m}\right\vert H|\varphi_{\mathbf{R}^{\prime}n^{\prime}%
}\right\rangle \right]  .
\end{align}
The terms containing $\mathbf{R}^{\prime\prime}$can be shown to vanish because
$\mathbf{R}^{\prime\prime}\times\mathbf{R}^{\prime\prime}=0$ and for every
$\mathbf{R}^{\prime\prime}$ there is $-\mathbf{R}^{\prime\prime}$. We then
obtain, writing explicitly the sum over $\mathbf{R}^{\prime\prime}$,%

\begin{align}
\mathbf{M}^{I}  &  =-i\sum_{\mathbf{R}^{\prime\prime}}G_{\mathbf{R}^{\prime
}n^{\prime},\mathbf{R}n}^{+}\mathbf{R}\times\nonumber\\
&  i\left[  \left\langle \varphi_{\mathbf{R}n}|H\left\vert \varphi
_{\mathbf{0}m}\right\rangle \left\langle \varphi_{\mathbf{0}m}\right\vert
\mathbf{r}|\varphi_{\mathbf{R}^{\prime}n^{\prime}}\right\rangle \right.
\nonumber\\
&  -\left.  \left\langle \varphi_{\mathbf{R}n}|\mathbf{r}\left\vert
\varphi_{\mathbf{0}m}\right\rangle \left\langle \varphi_{\mathbf{0}%
m}\right\vert H|\varphi_{\mathbf{R}^{\prime}n^{\prime}}\right\rangle \right]
.
\end{align}
As can be seen, the expression under the summation over $\mathbf{R}%
^{\prime\prime}$does not depend on $\mathbf{R}^{\prime\prime}$ so that the sum
may be replaced by the number of sites $N$. The itinerant part of the orbital
moment per unit volume is then%

\begin{align}
\mathbf{m}^{I}  &  =\frac{1}{\Omega}G_{\mathbf{R}^{\prime}n^{\prime
},\mathbf{R}n}^{+}\mathbf{R}\nonumber\\
&  \times\mathbf{[}\left\langle \varphi_{\mathbf{R}n}|H\left\vert
\varphi_{\mathbf{0}m}\right\rangle \left\langle \varphi_{\mathbf{0}%
m}\right\vert \mathbf{r}|\varphi_{\mathbf{R}^{\prime}n^{\prime}}\right\rangle
\nonumber\\
&  \text{ \ \ \ \ \ \ }-\left\langle \varphi_{\mathbf{R}n}|\mathbf{r}%
\left\vert \varphi_{\mathbf{0}m}\right\rangle \left\langle \varphi
_{\mathbf{0}m}\right\vert H|\varphi_{\mathbf{R}^{\prime}n^{\prime}%
}\right\rangle ], \label{mI}%
\end{align}
Using Eq. (\ref{rphi}) we find%

\begin{equation}
\left\langle \varphi_{\mathbf{0}m}|r_{k}|\varphi_{\mathbf{R}n^{\prime}%
}\right\rangle =-i\frac{\Omega}{(2\pi)^{3}}\int d^{3}qe^{-i\mathbf{q\cdot R}%
}\left\langle \frac{\partial\tilde{u}_{\mathbf{q}m}}{\partial q_{k}}|\tilde
{u}_{\mathbf{q}n^{\prime}}\right\rangle ,
\end{equation}
where we have used the fact that $\tilde{u}_{\mathbf{q}n}$ is periodic. The
matrix element of the Hamiltonian can be written as%

\begin{equation}
\left\langle \varphi_{\mathbf{R}n}|H|\varphi_{\mathbf{0}m}\right\rangle
=\frac{\Omega}{(2\pi)^{3}}\int_{BZ}d^{3}qe^{i\mathbf{q\cdot R}}\tilde{E}%
_{nm}(\mathbf{q}),
\end{equation}

\begin{equation}
\tilde{E}_{nm}(\mathbf{q})=\sum_{m^{\prime}}S_{nm^{\prime}}^{\dagger
}(\mathbf{q)}E_{\mathbf{q}m^{\prime}}S_{m^{\prime}m}(\mathbf{q}).
\end{equation}
Using the above two results and using the formula $\sum_{\mathbf{R}}%
\exp(i\mathbf{q\cdot R})=\delta(\mathbf{q})(2\pi)^{3}/\Omega$, as well as
recognizing the fact that for a periodic system $G_{\mathbf{R}^{\prime
}n^{\prime},\mathbf{R}n}^{+}$ depends on $\mathbf{R}^{\prime}\mathbf{-R}$, the
final result for the itinerant contribution is%

\begin{align}
m_{i}^{I}  &  =\epsilon_{ijk}\int_{BZ}\frac{d^{3}q}{(2\pi)^{3}}G_{n^{\prime}%
n}^{+}(\mathbf{q})\nonumber\\
&  \left[  \left\langle \frac{\partial\tilde{u}_{\mathbf{q}n}}{\partial q_{j}%
}|\frac{\partial\tilde{u}_{\mathbf{q}m}}{\partial q_{k}}\right\rangle
\tilde{E}_{mn^{\prime}}(\mathbf{q})+\frac{\partial\tilde{E}_{nm}(\mathbf{q}%
)}{\partial q_{j}}\left\langle \frac{\partial\tilde{u}_{\mathbf{q}m}}{\partial
q_{k}}|\tilde{u}_{\mathbf{q}n^{\prime}}\right\rangle \right]  \label{mit}%
\end{align}
This together with Eq. (\ref{mloc}) constitute our main results. Unlike the
non-interacting case with no band crossings for which $G$ is diagonal in the
band index, the summation over the band indices involves both intra- and
inter-band matrix elements, reflecting the possibility of mixing between
different bands when the interaction among the electrons is switched on or
when there are band crossings. It is noteworthy that from perturbation theory
we have \cite{ceresoli2006}%

\begin{equation}
\left\vert \frac{\partial u_{\mathbf{q}n}}{\partial q_{i}}\right\rangle
=\sum_{m\neq n}\frac{\left\vert u_{\mathbf{q}m}\right\rangle \left\langle
u_{\mathbf{q}m}\right\vert v_{i}\left\vert u_{\mathbf{q}n}\right\rangle
}{E_{\mathbf{k}m}-E_{\mathbf{k}n}}%
\end{equation}
so that $\left\langle \frac{\partial u_{\mathbf{q}n}}{\partial q_{i}%
}|u_{\mathbf{q}n^{\prime}}\right\rangle =0$ if $n^{\prime}=n$.

Our formula for the orbital moment factorizes into two parts: the one-particle
part, expressed by $u_{\mathbf{q}n}$ and $E_{\mathbf{q}n}$, contains
information about the underlying band structure of the system whereas the
interacting part, carried by the Green function, contains information about
exchange and correlations of the interacting electrons.:%

\begin{equation}
G^{-1}=G_{0}^{-1}-\Sigma\label{G-1}%
\end{equation}
where $\Sigma$ is the self-energy and $G_{0}$ is a non-interacting Green
function expressed in terms of $u_{\mathbf{q}n}$ and $E_{\mathbf{q}n}$. The
formula is quite general and provides a convenient means of including the
effects of exchange and correlations through the many-electron Green function.
It goes over into the non-interacting formula as the self-energy is switched
off. As pointed out earlier, our derived formula for the orbital magnetic
moment is only valid for systems with zero Chern number, where the Bloch
orbitals always can be choosen to be smooth and obey the periodic gauge
$\psi_{\mathbf{k}n}=\psi_{\mathbf{k+G,}n}$. However, when the Chern number
becomes nonzero, the period gauge is no longer valid. As a consequence
integrals around the perimeter of the Brillouin zone will not vanish; there
will be contributions due to singularities located in the interior of the
Brillouin zone \cite{thonhauser2006}.

Let us now apply our formula to a non-interacting insulating system without
band crossing in which case the transformation matrix $S=1$. In this case $G$
is diagonal in the band index and $\left\langle \frac{\partial u_{\mathbf{q}%
n}}{\partial q_{i}}|u_{\mathbf{q}n^{\prime}}\right\rangle =0$ if $n^{\prime
}=n$ and we obtain%

\begin{align}
m_{i} &  =-\epsilon_{ijk}\int\frac{d^{3}q}{(2\pi)^{3}}n_{F}(E_{\mathbf{q}%
n})\nonumber\\
&  \times\operatorname{Im}\left\langle \frac{\partial u_{\mathbf{q}n}%
}{\partial q_{j}}\right\vert H(\mathbf{q})+E_{\mathbf{q}n}\left\vert
\frac{\partial u_{\mathbf{q}n}}{\partial q_{k}}\right\rangle ,
\end{align}
multiplying by $-e/2c$ (Gaussian units) to obtain the magnetic moment. This
result is in agreement with that in Refs.
\cite{thonhauser2005,ceresoli2006,xiao2005}. We have also applied this formula
to the Haldane model \cite{haldane1988} and found good numerical agreement
with previous calculations \cite{thonhauser2005,ceresoli2006,xiao2005}. When
there are band crossings, $S$ may not be a unit matrix anymore and
consequently the formulas in Eqs. (\ref{mloc}) and (\ref{mit}) may yield
additional contribution not included in the previous results.

In summary, we have derived a general formula for the orbital magnetic moment
of interacting electrons in solids in which the effects of exchange and
correlations are embodied in the many-electron Green function. The derivation
does not rely on the presence of surface but simply makes use of the
translational property of the lattice. By expanding the Green function in
Wannier orbitals constructed from the underlying one-particle Bloch states of
the system a seamless connection with the non-interacting formula is achieved.
The Wannier orbitals are used only as a tool and the final formula is
expressed entirely in terms of the Bloch states and the many-electron Green
function written in the Bloch basis. The use of the Green function allows for
a transparent physical interpretation of the formula. The formula provides a
convenient access to inclusion of exchange and correlations beyond the
commonly used local density approximation. For example, the self-energy needed
in Eq. (\ref{G-1}) can be calculated using the standard \emph{GW}
approximation for weakly to moderately correlated materials and the LDA+DMFT
or the recently developed \emph{GW}+DMFT \cite{biermann03}, in the case of
systems with strong onsite correlations.

\begin{acknowledgments}
We would like to thank K. Terakura for valuable comments and discussions
during the course of this work. This work was partly supported by the Swedish
Research Council.
\end{acknowledgments}

\end{document}